# General Students' Misconceptions
# Related to
# Electricity and Magnetism

Cristian Raduta
Physics Department
The Ohio State University

**Advisor: Prof. Gordon Aubrecht**

# Table of Contents

5) Examples of methods from the literature that have been used to address these incorrect or insufficient ideas

6) Possible explanation of student ideas using p-prims

7) Conclusions

## *1) Introduction*

Electromagnetism, being much less intuitive than mechanics (where a lot of sources of misconceptions have been documented), has in addition to the common sources of misconceptions borrowed from mechanics other sources related to the abstract new concepts of electric and magnetic fields. Some Physics Education research intended to detect the major areas of misconceptions in the field of magnetism and electromagnetic related phenomena has been done.

This paper has several purposes: to

a)  give an overview of the major areas of misconceptions covered in the literature;

b)  suggest other areas of misconceptions not covered in the literature, and

c)  suggest possible reasons for these misconceptions;

d)  speak about the methodologies of the research that have been focused on identifying these problems of understanding, and also to give some examples from the literature that have been used to address these incorrect or insufficient ideas;

e)  determine what, if anything, there is in common about the insufficient ideas students have about magnetic concepts;  and

f)  discuss whether a p-prim approach or some other way would be a better way to think of these results.

## 1) Misconceptions related to electricity and magnetism

Students have a lot of misconceptions about Physics even in the field of mechanics, which is much more intuitive and understandable than the field of electricity and magnetism. First, I am going to present some of the research that has been done to address some of the most important misconceptions related to E&M.

### A. *Students' misconceptions related to applying Faraday's law*

For example, many students have difficulties in understanding the induced emf and how it is produced. Let's take the circuits shown below:

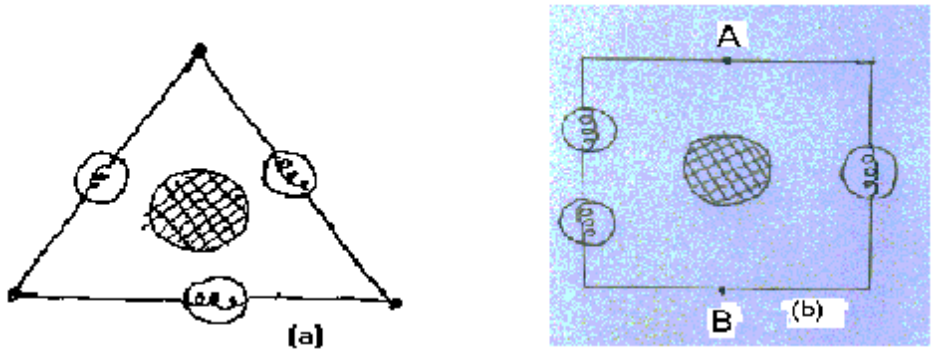

Fig.1. Equivalent three-bulb circuits; in a) the bulbs are arranged symmetrical about the source of emf; in b) the bulbs are arranged asymmetrically.

P.C. Peter (Ref. 1) has shown that although these are entirely equivalent circuits, many students state that the two bulbs on the left side of the asymmetric circuit will be dimmer, reasoning incorrectly that the emf on the left side, $\xi/2$, drives two bulbs in series, while the emf on the right side, also $\xi/2$, drives only one bulb.

The right answer, that all the bulbs are equally bright, requires an understanding of the fact that the total induced emf drives the bulbs in series, however they are geometrically placed around the solenoid, and whatever the shape of the circuit (triangular or rectangular, etc.) He showed also that, just using batteries, bulbs and a solenoid, we could create an almost endless number of problems that could uncover other students' misconceptions in this field.  For example, let's take the circuits shown below:

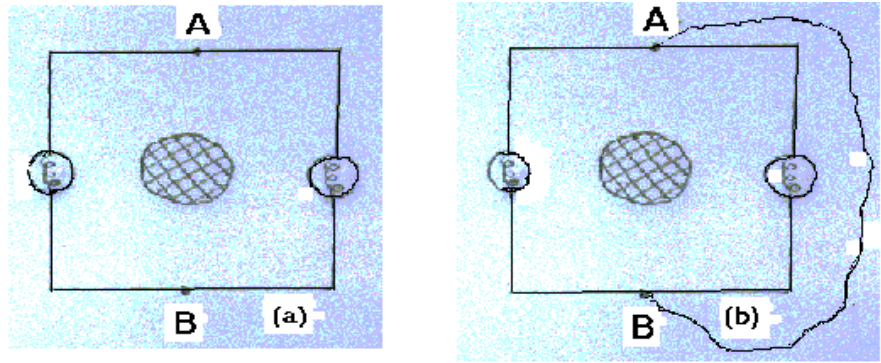

Fig 2. a) two-bulb two circuit with a solenoid that generates a constant emf;
b)  two-bulb circuit shorted around the right side.

Asking the students how the brightness of each bulb changes after connecting a wire between A and B as shown above, Peter (Ref.1) noticed a lot of confusion among the students. In general these kind of two-loop circuits in which we have in addition an induced emf are very confusing for the students. The important point here, as observed by Romer (Ref. 2), is that the topology of the circuit is very important when there are induced emf's, unlike the case of ordinary dc circuits, which can be deformed in any manner, as long as the ordering of the elements remain the same.

Another misconception that was documented by Bagno's study (Ref.6) is that students have difficulty in determining the direction of the induced emf (actually the "-" sign from Lenz law: $d\Phi/dt = -\xi$). It is suggested that the major source of difficulty has to do with fuzzy encoding. An examination of the relevant textbooks suggests that sentences such that "*the induced current resists its cause*" are too vague. Students could very easily interpret these sentences incorrectly. For example, "*oppose the change*" could be easily interpreted as meaning "*being in the opposite direction*".

**B. *Misconceptions related to the interaction between the magnetic field and electric charges***

Students' difficulties in understanding the interactions of electric charges with magnetic fields have been documented by Maloney (Ref. 3). He suggested that this may be caused, at least in part by an alternate idea (a p-prim). Giving the students the figure below, and requiring the students to rank these situations from the strongest attraction to the strongest repulsion, on the basis of the force exerted by the magnet on the charge, he found several interesting misconceptions.

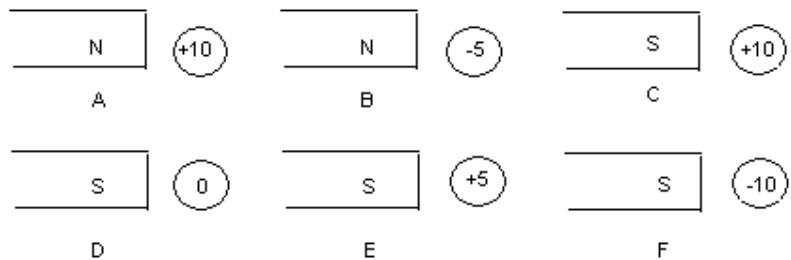

Fig. 3. Different electric charges placed close to the north and south magnetic poles of some permanent magnets

Most of the students answered this survey taking the N pole as being, or acting as though it were, positively charged. He implies that a lot of students think of magnetic pole as being electrically charged: "*Magnet of opposite charge will pull electrons.*" A second category of the students surveyed by Maloney, spoke of the effect of the poles as attractive or repulsive, but they made no statement about the poles' charge. An example in this sense is the answers of most students related to the figure from below: "*a negative charge will be attracted to the N pole and a positive charge to*

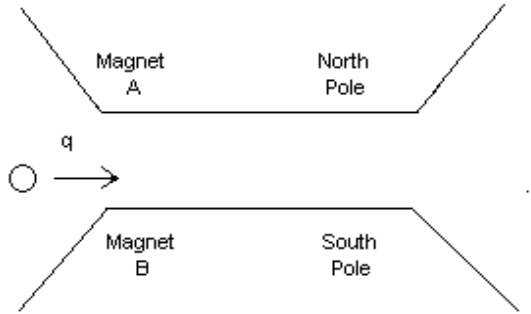

the S pole."

Fig 4. An electric charge coming with velocity between the north and south poles of a permanent magnet.

The tendency of students to calculate the magnetic force for situations where the charges are not moving, or are moving parallel or anti-parallel to the field, is known to anyone who has taught this topic.

Another misconception in this sense is that magnetic poles exert forces on electric charges in the plane of the charge and magnet, regardless of whether the charge was moving or not (D. Maloney, Ref.3)). What is interesting to notice is that this misconception prevailed both before and after instruction of the E&M course. Even students who have not studied the topic, used strategies in answering the questions that can be inferred to be rooted in a "*magnetic poles are charged*" alternate idea.

### C. Students fail to recognize the important ideas from E&M

In a study done by Bagno and Eylon (Ref. 4), they asked the students to summarize in a few sentences qualitatively the main ideas of electromagnetism, according to their order of importance, and found the following interesting results. A high proportion of the students considered Ohm's law to be one of the most important ideas of electromagnetism, which is actually consistent with previous findings (5), which the author labeled humorously: "*The three principles of electromagnetism: V = IR, I = V/R, R = V/I.*" Also, the symmetry that exists between the electric and magnetic fields was not reflected in the student' summaries. Fewer than 5% of the students surveyed mentioned the production of magnetic field by a changing electric field (**∇xH = J + ∂D/∂t**-Maxwell's first equation; **ΔE/Δt→B**—Maxwell's second equation). Even though some of the students surveyed by Bagno remembered the correct formula, only 10% of the students who remembered the correct formula claimed that a change in the magnetic field is accompanied by an electric field. This is another proof of the fact that students do not relate the labels "Lenz's law" or "induced emf" to the production of an electric field.

In a survey done with university level students in France and Sweden (S. Raison et al., Ref. 16), difficulties arise out of two issues: (i) a causal interpretation of some relationships, (ii) the students' need for an effect, motion of some kind, to allow them to accept the existence of a field. Students accept the existence of a cause only when they can imagine an effect. In response to a question involving insulators, many students gave an argument that: "*charges cannot move in an insulator, therefore there is no electric field.*"

Also students interpret formulas as if the quantities mentioned to the right of equal sign were the cause of those mentioned to the left. In the case of Gauss's theorem, this suggests a response such as "*to calculate the electric field, I only need the internal charges,*" or "*the electric field is due only to the internal charges.*" But the students fail to say anything about the charges situated in the exterior of a symmetrically charged sphere (more than 80% of students' responses were attributed this kind of reasoning).

Their study also concluded that for most of the students, electrostatics and electric circuits are two unconnected subjects. A lot of students think that current is the cause of the field, reversing the cause and the effect. Rainson et al. ( Ref. 16) conclude that the above misconceptions ("*field if mobility*"-in order to accept the existence of a field the student needs to see a motion; "*cause in the formula*"-an erroneous interpretation of a mathematical relationship: the quantities on the right side of one relation are seen as the cause for the quantities from the left side) are determinant for the difficulties that students have with a very basic principle of physics, the superposition of electric and magnetic fields.

### D. Students see the electric and magnetic fields as having a "static" nature

An important misconception to notice is that many students consider the electric field to have a static nature, in the sense that the field exists in the space and applies forces on charges, and it does not change even when a new charged particle enters the region. Indeed, from an interview by Bagno et al, when the students were given the statement "*A charged particle enters a region with a constant electric field. The field in this area changes because of the new charge,*" 40% of the students answered incorrectly, from which 82% of them were saying that "*the electric field is a "property" of the region-its task is to apply force on a charge in it.*" The authors (Bagno et al., Ref. 4) explain this misconception by reference to the presentations of the most textbooks, which support this perception of the students, since the electric field, a difficult and non-intuitive concept, is presented merely as a force applier. They also notice that in general problems from the textbooks deal with static situations such as: "*four charges are fixed in the four corners of a rectangle; find the resultant electric field,*" and do not illustrate the dynamic nature of the electric field. Even in the problems in which charged particles are entering a region with a constant electric field, students are almost never asked about the new field (they are usually asked about the path of the particle, its velocity, etc). Chabay and Sherwood (Ref.6) have made an attempt to develop a dynamic conception of electric fields in their recent instructional materials (they included and emphasized also problems in which the students are required to find the new electric field after an electric charge entered into a region with a constant electric field).

### E. Misconceptions related to erroneous interpretation of a symbol or due to ambiguous presentations from the textbooks

When the students surveyed by Bogno et al. were asked whether the statement "*at the point where the electric field is zero, the electric potential is also zero*" is true or not, 62% of the students chose incorrect answers. The authors offer several explanations for this. It seems that many students don't differentiate between concepts of potential and potential difference (student reasoning: $\mathbf{E} = 0$, V(voltage) = $\int \mathbf{E} \cdot d\mathbf{r} = 0$, P(potential) = 0). Another source of this misconception could be an erroneous interpretation of a symbol. As usual, textbooks are also a source for misconceptions. The authors explain that the presentations from textbooks suggest the possibility that the proximity of introducing the electric field and electric potential, as well as the similarity of the formulas of their calculation, may lead to the confusion of the terms. They say that the problems from textbooks lead to the same impression.

S. Tornkvist et. al. (Ref. 18) in one of their interviews asked the students to draw the field lines that can account for a given force vector in a given point (see Fig. 5).

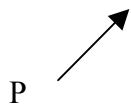

P

Only 13% of the students considered an inhomogeneous field as the answer to this question, although they have been given such fields in previous questions. 79% of the students drew straight equidistant field lines. The authors think that an explanation for this could be the heavy emphasis in textbooks on the homogeneous electric field between two parallel capacitor plates.

Harrington in one of his papers (Ref. 19) found that a lot of students (28% of the interviewed students) gave answers to his E&M related questions consistent with the idea that an object that is neutral can be considered negatively charged. Indeed, one of his students stated: "*It is negative charge because it is not charged. Isn't that what negative means?*" Another student stated: "*Doesn't positive mean yes, and negative means no?*" These misconceptions could be explained also by the textbooks, which don't repeatedly emphasize the distinction among positive, negative and neutral charges.

*F. Misconceptions related to the direction of the Lorentz force and to the application of the right hand rule*

Giving the students the statement, "*the velocity of a charged particle moving in a magnetic field is always perpendicular to the direction of the field,*" 37% of the students agreed, out of which group (those students who answered incorrectly) 81% gave answers similar to: "**v**, **B***, and* **F** *are always perpendicular to each other according to the left hand or right screw law.*" The authors suggest that the difficulty of the students with this statement is caused by the fact that most of the problems in electromagnetism deal with charged particles whose initial direction is perpendicular to the direction of the magnetic field. This may lead the students to incorrect generalization that the path of a charged particle in a magnetic field is always circular. And, indeed, this is what is happening, because, when asked whether the statement "*the path of a charged particle moving in a magnetic field is circular*" is true or false, 60% of the students considered it to be correct.

### G. E&M misconceptions related to mechanics misconceptions

The same authors (Bogno et al) showed that misconceptions in one field (mechanics) may cause difficulties in another (magnetism). When they asked the students whether or not the statement "*a constant magnetic field never changes the speed (magnitude of velocity) of a charged particle which moves in it*" is true, 46% answered it incorrectly. Forty percent of the students who gave incorrect answers, attached acceleration only to a change in the magnitude of velocity and not in its direction, a well documented misconception in mechanics.

Galili (7) studied students' misconceptions from E&M related to the well known mechanical misconceptions, thus demonstrating their persistence on the one hand, and indicating the relevance of mechanical misconceptions beyond mechanics, on the other hand. The students were given the pictures from below and were asked to choose the correct answer (the relative angular locations of the suspended charges).

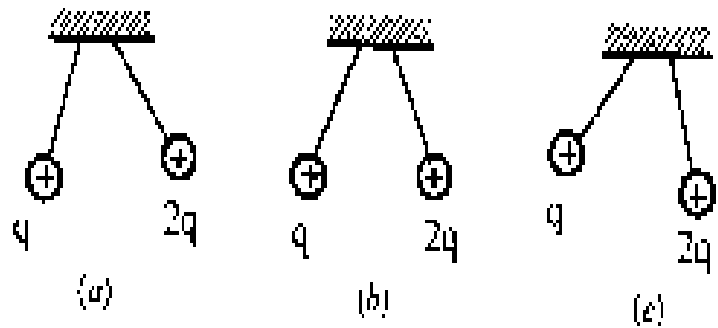

Fig.6. A double pendulum with two different charges of equal masses

The straightforward application of Newton's third law (the masses of the two charges are the same) could immediately have provided the correct answer (a symmetrical angular displacement). However, only a third of the students gave the right answer. They apparently considered an "electrical" question using only "electrical" tools, which some of them applied correctly, some not. Their reasoning which could sound like, "*the bigger the charge, the bigger the force*" (a p-prim), was wrongly considered, and would lead to the violation of a basic principle, supposed to have been mastered in mechanics: the symmetry of the force interaction (the action-reaction principle, the third Newton law).

Another two examples of students' failing to apply the principles of mechanics to the E&M problems are given below.

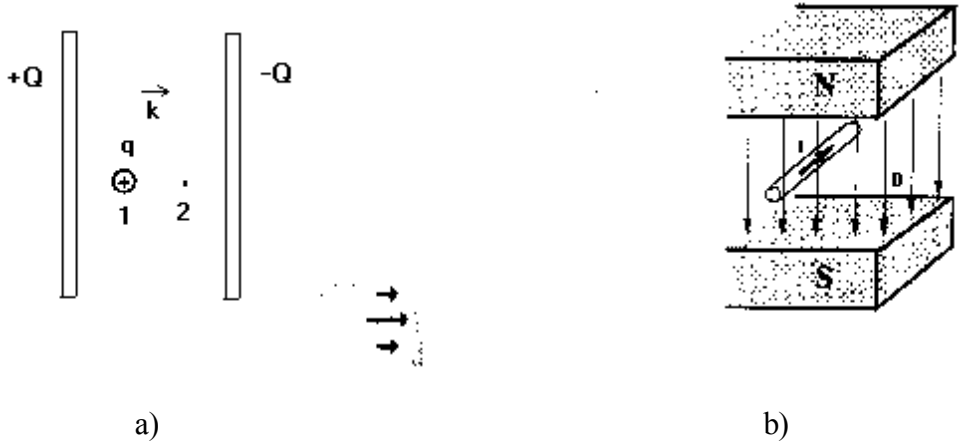

a)                                                    b)

Fig. 7.

The answers of the students show once again that they tend to miss the general considerations of action-reaction principle. Indeed, they seem to think in terms of "*field on charge action*" (indeed, 75% of the students answered in the way shown in figure 7(a)).

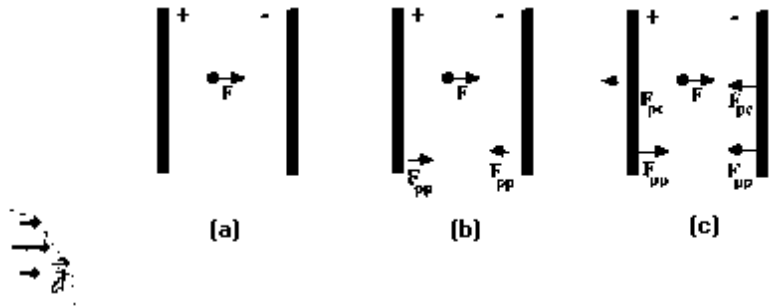

Fig. 8.

The same failure to apply Newton's third law was observed even more clearly to the next question given to the students associated with Fig.6 (b), which was actually very similar to the first question, only that it was formulated in the context of a magnetic field. Only about 3% of all the students showed a force applied to the magnet due to the current-carrying wire.

Work-energy considerations in the presence of electromagnetic fields represent another important aspect of students' understanding. Galili (Ref.7) surveys in this sense show that this is also a critical point. When students were asked about the sources of the kinetic energy increase of the electric charge placed in an electric field (Fig. 6 (a)), less than a third of the students answered correctly. But even among the students who answered correctly, few of them have gone beyond the general statement of *"energy transformation,"* which does not mean necessarily that that the students understand the process. So, Galili's research proves one more time the difficulty of the students have in including the concept of *"field"* within the mechanics framework previously acquired in the physics courses.

Student answers to another question given by Galili (Fig. 8) pointed out the difficulty that the students have when they have to deal simultaneously with both mechanics and E&M

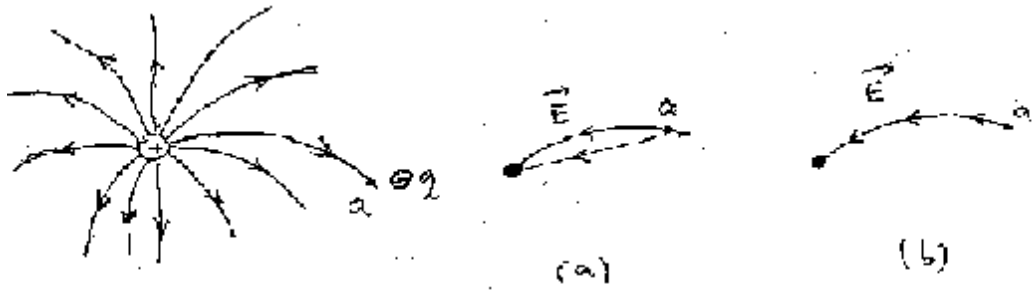

Fig.9. The students were asked to draw the trajectory of the negatively charged particle in the field created by a positive charge;

Only two of the students interviewed gave the correct answer. Most of the others answered the question either as in Fig.8 (a) or as in Fig.8 (b). They have the same well-documented mechanics misconception of confusing the trajectory with the force-line. The author explains these difficulties of the students by the change of the tools needed to treat the interaction, namely the introduction of a field concept. The field concept presents a topic of high conceptual difficulty for the students. It is commonly introduced through the formal operational definition and it could influence in a wrong way the understanding of other problematic general principles previously assimilated by students while studying mechanics (among them we discussed about Newton's third law, commonly referred the action and reactionlaw, and about the work-energy interrelations-which actually by themselves are complicated problems in learning mechanics-e.g. Brown and Clement 1987, Brown 1989, Lawson and McDermott 1987).

The same misconception of confusing the trajectory of a charged particle with the field lines was documented by Tornkvist et. al. (Ref.18). He asked the students to draw a likely trajectory for a particle with zero initial velocity in a given point in a given field

(see Fig. 10).

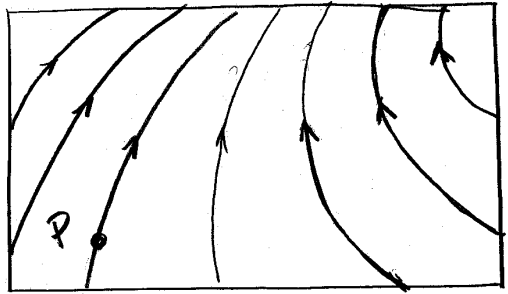

Fig. 10

76% of the students made the trajectory follow the field line. 7% drew it toward the supposed pole (assumed, implied one), and 6% were completely confused. Only 11% of the students offered reasonable trajectories.

### 3) Suggested areas of "misconceptions" related to magnetism and electromagnetism that may have been missed by the current research

Compared to other areas of physics such as mechanics, electricity, thermodynamics, etc, where we have a lot of results of Physics Education research available, the field of magnetism and electromagnetism has been much less explored in this sense. So, there should be a lot of areas of misconceptions not yet detected.

#### A. Mathematics related misconceptions

One very big source of misconceptions in electromagnetism is of course the mathematical tools involved, which are a little more sophisticated than those used, let's say, in mechanics. The students should be used to working on a regular basis with vector products, derivatives, gradients, etc. Let's take for example the Lorentz force:

$$\mathbf{F} = q\mathbf{v}\mathbf{x}\mathbf{B}; \quad F = qvB\sin(\mathbf{v}, \mathbf{B}) \tag{1}$$

Students have been taught in almost all the books to find the sense and direction of the Lorenz force using the "right hand rule." But this is very easy to forget: "*What hand should I use, right or left? But what if the charge is negative?*" The students should learn to use the vector product that will appear a lot in the field of electromagnetism. A common misconception that was noted earlier is that a lot of students think that in the Lorentz force expression, the velocity and the magnetic field must be perpendicular to each other. This is partly, as we said earlier, because most of the applications in magnetism deal with a charged particle coming into a region containing a magnetic field with the velocity being perpendicular to the magnetic field. But this is also because the students do not know how to handle the vector products. From relation (1), using the vector product, it is much clearer than using the right hand rule that the Lorentz force is perpendicular to the plane made by **v** and **B**, and also that the angle between **v** and **B** is not necessarily equal to 90 degrees.

Another mathematics source of misconceptions could be the scalar product which is involved in the calculation of the electric and magnetic flux. Let's take, for example Faraday's law:

$$\iint \mathbf{E} d\mathbf{r} = \xi = -d\phi/dt; \qquad \phi = \iint \mathbf{B} d\mathbf{S} \qquad (2)$$

As in the case where the students assumed that for the Lorentz force, the velocity and the magnetic field are perpendicular to each other, also here they could very easily implicitly assume that the magnetic flux density **B** is always perpendicular on the surface through which they calculate the magnetic flux; but situations such as the one from Fig.9 shows us that this is not always true.

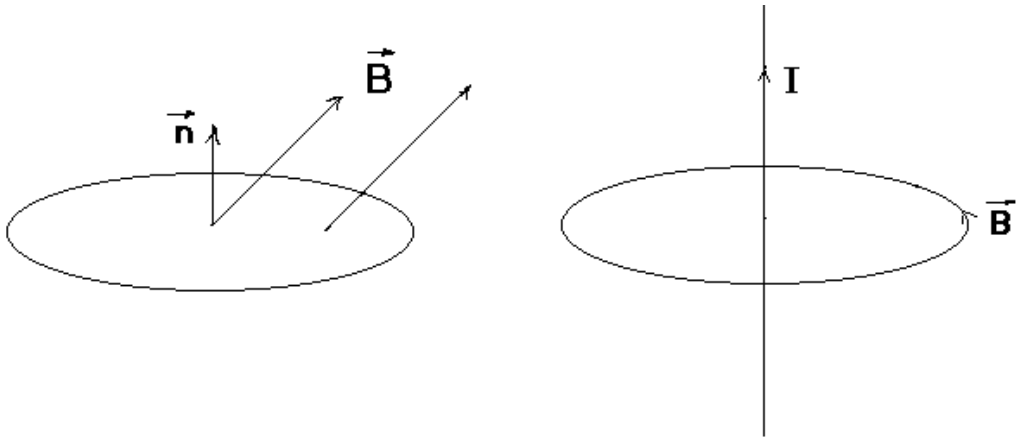

Fig.11

Such mathematics-related misconceptions could generate a lot of misconceptions in the field of magnetism and electromagnetism.

### B. Another possible source of misconceptions: the analogies between the electric and magnetic fields that students are tempted to make

Another big source of misconceptions in the field of magnetism could be given by the analogies the students are tempted to make between the electric and magnetic fields, analogies that are not always correct. For example (as in the misconception noted earlier in which the students were thinking that magnets could attract electric charges initially at rest), it is very easy for them to make the next connection: "*If the electric field is created by electric charges, than also the magnetic field should be created by magnetic charges.*" But from the Maxwell relation, div $\mathbf{B}$ = 0 (unlike div $\mathbf{D}$ = $\rho$) we know that we don't have magnetic charges (magnetic monopoles).

### C. Do the students make the connection between Maxwell' equations and the derived laws (Ampère's law, the Biot-Savart law, Faraday's law and Coulomb law for the electric forces)?

Other sources of misconceptions could be very easily the Ampère and the Biot-Savart laws:

$$\int \mathbf{H} \cdot d\mathbf{S} = I_{enclosed} \text{ (Ampère's law)}; \qquad d\mathbf{B} = \mu_0 I (d\mathbf{l} \mathbf{x} \mathbf{r})/4\pi r^3 \qquad (3)$$

One source of misconceptions here is of course the one generated by mathematics (we discussed this earlier). But, if we would ask the students let's say about the magnetic field created by a long straight current at a distance r, maybe a lot of students would come up with the correct magnetic field shape or sense, or even with the correct quantitative formula memorized like a poem. But I am sure that very few of them would make any connection between the shape and magnitude of the magnetic field that they memorize and the Ampère or Biot-Savart laws. If we would ask them further more how the Biot-Savart or Ampère laws were derived, from which of Maxwell's equations (1-st Maxwell law $\nabla \times \mathbf{H} = \mathbf{J} + \partial \mathbf{D} / \partial t$), I'm sure that we would create even bigger confusion.

We would create the same confusion if we would ask them about the relation between the second law of Maxwell ($\nabla \times \mathbf{E} = -\partial \mathbf{B} / \partial t$) and the Faraday's law ($\int \mathbf{E} \cdot d\mathbf{S} = -d\Phi/dt$), or the relation between the fourth law of Maxwell ($\nabla \cdot \mathbf{D} = \rho$) and the Coulomb law ($\mathbf{F}_{21} = q_1 q_2 \mathbf{r} / 4\pi\varepsilon_0 \ r^3$). It is very possible that they won't make any connection between these laws, considering them as independent.

### D. Shape of Lorentz force--another possible source of misconceptions

A big source of misconceptions in the field of magnetism, I think is the shape of the Lorentz force: $\mathbf{F} = q\mathbf{v} \times \mathbf{B}$. For the first time in their studies in Physics students encounter something totally different. Up to that moment, the forces that they have learned were always along the direction of the two objects (in mechanics), or, where it was also included a field (as in electrostatics or with the gravitational field), the forces were along the direction of the field. Now, the situation is different. The direction of the force is perpendicular to the magnetic field and the velocity of the particle, and so is something totally different from what they used to see.

Of course there are more areas of misconceptions related to electromagnetism that I haven't touched in this paper. The suggested areas of misconceptions that I gave above, of course, are at this moment just speculations (assuming that they have not already been taken into account by others). But by designing specific surveys focused on these aspects, we could see how serious these potential areas of misconceptions are.

## 4) Common things about the incorrect or insufficient ideas students have about magnetic concepts

### A. Analogies between the electric and magnetic fields are common to a lot of students' misconceptions

One thing in common about many of their misconceptions is the fact that they are tempted to make analogies between the electric and magnetic fields, analogies that, many times do not work properly. As we discussed earlier, they are tempted to think that magnets interact with static charges (the N pole attracts negative charges and the S polo attracts positive charges). They are also tempted to think that because we have electric charges, we should definitely have magnetic charges. They do not feel too comfortable with the direction of the Lorentz force, and they are tempted to think that it should be in the same plane with the magnetic field and the charge. Probably they don't feel too comfortable with the idea that the magnetic field is produced by moving electric charges, while in the case of electric field, the charges do not necessarily have to move in order to create an electric field.

### B. Mathematics related misconceptions generate many of the E&M related misconceptions

Another thing in common about their incorrect or insufficient ideas about the magnetic concepts is their poor understanding of the vector and scalar products. These products appear almost anywhere in the field of magnetism: in the Lorentz force, in the magnetic flux, in Ampère's law or in the Biot-Savart law. If they will not be able to be more confident with these simple products, they will have basic problems with a lot of concepts from magnetism. For example, an incorrect understanding of one magnetic concept, (such as the magnetic flux, which the student might get correct up to a cosine of an angle), could generate incorrect answers in a chain (for example, he will write Faraday's law correctly again up to a cosine). In the same way, if the student does not master sufficiently well the vector products, he (or she) will not apply correctly the Biot-Savart law, the Lorenz force or Maxwell's second equation ($\nabla$x$\mathbf{E}$ = -$\partial\mathbf{B}/\partial$t) from which is derived Faraday's law. So, these mathematics-related misconceptions will be reflected in E&M misconceptions.

### C. Textbooks are also a source of E&M related misconceptions

A big source of a lot of misconceptions that students have in the field of electromagnetism is the way that textbooks present the subject. As Bagno noticed, in most of the textbooks, Ohm's law is central, and this is happening in the presentation of the theory and also in the exercises associated to the theory. So, it is not surprising that Ohm's law was found in Bagno's surveys to be considered by students one of the most central laws of electromagnetism. Bagno et al also observed that most of the textbooks do not emphasize the idea that a change in the magnetic field is related to the production of an electric field, while the idea of induced emf is emphasized in the theory and the associated exercises. So, it is not surprising that students do not associate labels such as "*Lenz's law*" or "*induced emf*" with the production of the electric field.

### D. Mechanics-related misconceptions are reflected in some of the E&M-related misconceptions

Also, misconceptions that students have from mechanics could generate other misconceptions in the field of E&M (I gave several examples earlier). But we could have concepts from mechanics which the student seems to understand pretty well in the context of mechanics, that, when integrated in the context of E&M could create a lot of misconceptions.

For example in the problem associated with Fig.6—a double pendulum with two different charges of equal masses—it is likely that a high percentage of the students who gave a wrong answer to the question related to the angular displacement of the two pendula, would have answered this type of question—involving the 3-rd principle of Mechanics—correctly in the field of mechanics. It seems that for some of the students it's difficult to deal simultaneously with concepts from both fields (Mechanics and E&M).

## 5) Examples of methods from the literature that have been used to address these incorrect or insufficient ideas

Surveys and interviews are the most common way in which researchers in Physics Education try to detect the students' misunderstandings. For example, Maloney (Ref. 3), wanted to determine whether or not the students were thinking of magnetic poles as exerting forces directly on the electric charges, in a manner similar to the behavior of electrostatic charges. He gave the students two problem situations, presented in different formats. Both formats were designed so that he could determine the strategies the students applied to the problems.

One of the problem situations had an electric charge moving at a right angle to the field between the pole faces of a two permanent magnets, in the plane of the magnetic field. Changing the polarities of the two magnets and the sign of the charge, as in the table below, several similar problems could be given to the students.

Table 1.

| Problem type | Polarities | Strengths | Charge |
|---|---|---|---|
| 1 | Both N | Equal | + |
| 2 | Both N | Equal | - |
| 3 | Both S | Equal | + |
| 4 | Both S | Equal | - |
| 5 | Both N | Different | + |
| 6 | Both N | Different | - |
| 7 | Both S | Different | + |
| 8 | Both S | Different | - |
| 9 | Opposite | Equal | + |
| 10 | Opposite | Equal | - |

 By analyzing their answers, the author could detect the strategies used by students . The student's sequence and the sequence of an identified strategy were considered to match when there were no more than three differences between the two sequences. The students were asked the same question for all the similar problems: "Which way will the electric charge be pushed as it moves through the area between the poles? " (the possible answers were that "*the charge will move toward magnet A,*" or "*the charge will move toward magnet B,*" or "*the charge will go straight from this view*" –see Fig. 4). The most popular individual strategy was the one that took the N-pole as being, or acting as though it were, positively charged. Discounting the people who were not consistent (the muddlers), the next most popular strategy had both types of charges being attracted to the stronger magnetic pole, whatever its polarity. The results of the survey could be seen in Table 2.

Table 2. Percentages of students using each rule for particles in a magnetic field

| Rule | Class of Spr 84 | Class of Sum 84 |
|---|---|---|
| + to N, - to S | 8 | 14 |
| - to N, + to S | 28 | 29 |
| + and – both to stronger (weaker) | 21 | 19 |
| Muddle | 19 | 14 |
| Miscellaneous | 14 | 14 |
| All equal | 7 | 10 |
| No fit | 4 | 0 |

The second problem situation (the "ranking task") that Maloney gave to the students had an electric charge sitting at rest near one pole of a permanent magnet. (see Fig3). The students were asked to rank the interactions between the magnets and the charges from the strongest to the weakest. He gave the "ranking task" problem to four classes, among which two hadn't had any college level instruction on electromagnetism. The results of the survey were summarized in the table 3 from below:

Table 3. Percentage of students using specified rule on ranking task.

| Class | N attr - | N attr + | All same | Misc | No fit | n |
|---|---|---|---|---|---|---|
| 202 Spr 84 | 50 | 14 | 0 | 12 | 24 | 58 |
| 202 Sum 84 | 59 | 27 | 0 | 0 | 14 | 22 |
| 204 Spr 84 | 42 | 13 | 11 | 7 | 26 | 114 |
| 204 Sum 84 | 55 | 20 | 15 | 0 | 10 | 20 |

What is important to notice here, is the similarity between the pre-instruction and post-instruction results. We notice also the similarities within the pre-test and post-test groups. For all classes the strategy that treats the N pole as being positive is the most popular. Analyzing all these results, Maloney could come up with all the suggestions that we've discussed earlier in the section of misconceptions.

Peter, trying to detect the students' misconceptions related to the double-loop circuits, in which there is also an induced emf due to a solenoid, gave to the students the next problem (see also Fig. 10.)

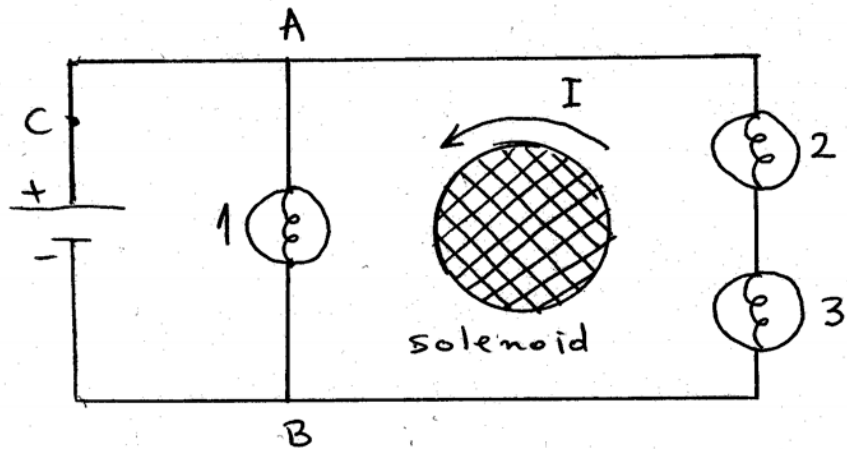

Fig.12.

In the circuit shown in the diagram, the current in the solenoid ( a long solenoid oriented perpendicular to the figure) is in the direction shown and is increasing linearly with time. The battery has a voltage equal to the emf of a loop around the solenoid. The bulbs are all the same, and for calibration, one bulb connected to the battery has brightness B1, two bulbs in series connected to the battery have brightness B2, three B3, etc. The internal resistance of the battery is much less than the resistance of the bulbs.

a)  Give the brightness of each bulb for the circuit as shown.

For each of the following parts, give the brightness of each bulb when the indicated alteration is made on the original circuit:

b) Bulb #1 is unscrewed.

c) Bulb #2 is unscrewed.

d) The wire is cut at C.

e) The circuit is shorted from A to B by a wire around the left side of the circuit.

f) The circuit is shorted from A to B by a wire around the left side of the circuit.

Out of fifty students answering the questions, the percentages of correct answers for each part were: a) 13%, b) 46%, c) 69%, d) 81%, e) 26% and f) 61%. The author explains that question a) had the lowest rate of success also because it requires looking at two different loops in order to obtain the answer. Also, questions b), c) and d) being simple one-loop circuits have a higher rate of success.

Bagno in one of her papers (4) uses a diagnostic survey that looked into students' knowledge representation in the domain of electromagnetism. Her investigation examined three questions:

1) Which ideas students view as central in electromagnetism? Are their key relationships summarized by Maxwell's equations?

2) Under what form do students represent the main ideas? Do they give also a qualitatively representation of the relations or only a mathematical one?

3) Do the students understand the key relations in electromagnetism? Do they know to apply the relations in solving problems?

Having these three questions as a central point of the investigation, Bagno designed the survey meant to detect the students' main ideas in electromagnetism as follows:

a) ***Free recall***: "*Summarize in a few sentences the main ideas of electromagnetism according to the order of their importance. Don't use formulas!*"

b) ***Cued recall***:  The cues were labels, intended to facilitate access. The task sounded in the following manner: "*Next to each of the following concepts, write as many relationships as possible that include the concept: (i) Electric Field (ii) Magnetic Field.*"

In order to test the form of representation for the electromagnetism key relationships, each statement was categorized into one of the following:

1) A qualitative verbal statement about a relationship or property of a concept. For example: "*An electric charge produces an electric field.*"

2) A verbal translation of a formula. Example: "*Current equals charge over time.*"

3) A mathematical formula: Example: $\mathbf{F} = q\mathbf{v} \times \mathbf{B}$.

4) A label. Example: "*Gauss's law,*" "*electric field.*" See Table 4.

Table 4.
Average performance of the various categories of form in the diagnostic study (N=250)

| Form | % out of total number of statements |
| --- | --- |
| (a) qualitative | 45% |
| (b) "verbal" formula | 20% |
| (c) formula | 0% |
| (d) label | 18% |

To test the conceptual understanding related to electromagnetism, Bagno gave her students several statements (e.g., "*A constant magnetic field never changes the speed (magnitude of velocity) of a charged particle which moves in it.*") for which the students had to answer whether or not they are correct. I analyzed several of those statements and students' answers in the section related to students' misconceptions to electromagnetism.

# 6) *Possible explanation of students' ideas using p-prims*

P-prims or "phenomenological primitives" could be another explanation for some of the students' misconceptions related to E&M. P-prims are relatively minimal abstractions (diSessa, (9)) of simple common phenomena. They are explanations that the students used to explain all kind of phenomena from the surroundings before learning any physics at all. Physics-naive students have a large collection of these p-prims in terms of which they see the world and to which they appeal as self-contained explanations for what they see. In the process of learning physics, some of these p-prims cease being primitives (and are seen by being explained by other notions), and some may even cease being recognized at all. Some of these p-prims could be the cause of some of students' misconceptions in E&M. In the table below we have a list of a few well-known p-prims.

Table5. A list of some well-known p-prims

| | |
|---|---|
| 1) Ohm's p-prims | --it comprises of *three elements*: **impetus, a resistance and a result;**<br>--qualitative correlations: increase in impetus implies an increase in result; increase in resistance means a decrease in result; etc<br>--very commonly used, high priority p-prim<br>--context of application: ex; pushing harder in order to make objects move faster<br>--Ohm's P-prim becomes profitably involved with the physical Ohm's Law as a model of causality and qualitative relations compatible with it;<br>--I think the definition want to look more savant than it really is. I would condense better this definition in *: "More is more, or bigger cause implies bigger effect"* |
| 2) Rolling and Pivoting | --a p-prim?  I think is to narrow the spectrum of contexts in which this could be applied<br>--rolling and pivoting, especially the latter one, are sometimes confusing for students |
| 3) Dying Away | Aristotel explicitly cited the dying away of certain actions like the dying away of sound from a bell as a primitive element of analysis that one does not seek to explain but simply is so;<br>Context of application: students assume a constant force is needed to maintain a constant velocity; |
| 4) Force as a mover (false intuition) | --force causes motion in the direction of the force, ignoring the effect of the previous motion; |
| 5) cause as a center, as a nucleus (not confirmed yet) | --students see the causes that produces a lot of effects as being condensed in a center, in a nucleus, like the Sun, electric charges, etc. |

For instance, in the article by Maloney (Ref.3), we have seen that before and after the instruction, most of the students thought that magnets interact with electric charges at rest; most of them took the N-pole as being, or acting as though it were, positively charged, etc. This is definitely an example of a p-prim, which was so strong rooted into the students' minds, that even after the E&M instruction it couldn't be eliminated. This general p-prim could be written as: "*opposites attract and likes repel each other.*"

The terms "magnetism" and "magnetization" were heard by the students even before they came for the first time in contact with physics or E&M, in contexts totally not related with electromagnetism. A lot of times, we've all heard expressions such as: "*this actor has a certain magnetism, a certain charisma*", or "*I feel magnetized by her or by him*". So, even before taken the course of E&M most of the students had their own vague idea (explained by a p-prim) about magnetism, or magnetization. Making also the analogy with what they see happening between a magnet and a piece of metal (or between the cinema stars and the fans-attraction), they are tempted to infer that the same happens between magnets and static electric charges. Also, if we analyze the expressions above, we can conclude that is very easy for the students to make the connection that it has to be a source for the magnetic field (the actor, or the man or woman in our example) similar to the one for the electric field, like some point or some place, which emits the magnetic field. Also, the lines of the field that they are inclined to think about should be radial, from the actor to the fans, like the rays of the Sun (from the Sun to Earth, from the magnetic charge to the electric charge). This is probably why they are also tempted to think even after the instruction that it should be a magnetic charge, which is responsible for the magnetic field. This could also explain why they don't feel comfortable with the shape of the Lorenz force, which is not "*along the ray*" as they would like to think about it. So, these are the first intuitive images that the students are tempted to think about, and which came perhaps naturally to most of us in our imagination when we wanted to heuristic explain the term of "magnetization" before learning E&M. All these things could be explained by general p-prims like: "*every effect (thing) should have a clear cause (source)*"—this is why the students think that it should be some kind of magnetic charge responsible for the magnetic field. Another p-prim could be: "*The interaction between things is happening in a radial way.*" This is why they don't feel comfortable with the shape of Lorentz force. Still another p-prim could be: "*cause as a center, as a nucleus.*" Students see the cause that produces a lot of effects, as being condensed in a center, in a "nucleus," like the Sun, electric charges, etc.

Another example of a p-prim emerges from the survey done by Galili (Ref. 7-see Fig. 5) Their reasoning that could sound like, "*the bigger the charge, the bigger the force*" is definitely a p-prim. They totally forgot about the basic principle of Mechanics (Newton's third law), or about the symmetry of the electric force of Coulomb. Their strong idea that in general "*something bigger should cause something also bigger,*" (in our case a bigger force and a bigger angle), lead them to forget about whatever other principles that they have learned along their physics studies. Something that is rooted for long time in their thinking structure, in their own way of seeing the things, of course that should be stronger than some rules (physics laws) that they have learned by heart, without having any resonance with their way of feeling and thinking. This p-prim that could have caused this kind of reasoning could sound like: "*more implies more, or bigger cause implies bigger effect.*"

It is important to know about a naive physics student repertoire of p-prims related to E&M. Knowing their p-prims, we could make all kind of analogies related to their p-prims when trying to teach them more advanced physics concepts.

# 7) Conclusions

Even though, not so much Physics Education research has been done in the field of magnetism and electromagnetism, many important students' misconceptions have already been detected. A lot of them appear to have their roots in the textbook presentations of the subject. Sometimes, things that should have been emphasized and stated several times in order to insure the successful understanding of the concepts, are just written very briefly in a way that tempts students to think that are not worth learning (or even remembering).

Other times the textbooks wording is very ambiguous, with double meanings possible, leading the students to understand something else than they should understand. Also the examination, by myself, of the textbooks shows that there is no emphasis on the qualitative analysis and verbal statement of relationships. Also, although some of the textbooks attempt to organize the information locally (e.g., within a single chapter) by making a summary or a table, there are no comprehensive attempts to organize the information at a global level (which would facilitate students connections between different concepts of electromagnetism-instead of seeing just the trees, to see the whole forest, relating the trees from the south with the ones from the north, and so on).

Another big source for students' misconceptions is also the mathematical tools that they must learn to handle while they are learning the subject of electromagnetism. A lot of students have difficulties with vector or scalar products, which are fundamental for the successful understanding of magnetism and thus electromagnetism. They probably have even more difficulties with gradients, divergences or laplacians.

Previous misconceptions from mechanics or the analogies that students make between the electric and magnetic field are other important sources of misconceptions. And why not admit that one of the biggest sources of misconceptions is the small amount of time allotted (and most of the time in a rush with the eyes on the watch) by the students to learning electromagnetism (and in general in college). Probably, the rush for getting the credits, for making so many other assignments for other classes that they take, for getting their degree done one way or another, is killing their real interest for deeper study of electromagnetism's concepts.

Maxwell's equations (and, in general the whole field of E&M) are constructed out of deep concepts that have been developed after a lot of thought done by brilliant Physicists along the history of Physics. That's probably why E&M cannot be totally successfully covered in a few hours per week, done in a rush by most of the students just for getting done the homework.

Probably we should do something to awaken their real interest for physics, for discovering the unknown in general, for wanting to probe deeper into the concepts of magnetism. Maybe a redesigning of the way the classes are taught, and the way home-works are treated would be a good first step. Maybe more analogies with things that they understand (and probably like) would be another idea. Or maybe raising the level at which they study magnetism in high-school a bit (for example, see Bagno (Ref. 4)- the level at which electromagnetism is done in high-schools in Israel is similar to the one done in the first year in US colleges; and this is happening in a lot of other countries-most countries from Europe) would be another idea. The difference between the level at which the magnetism was done in high-school and the one from the colleges is maybe to big to be so easily and fast assimilated by the students in such a short amount of time.

**Acknowledgement**:  I would like to thank Prof. Gordon Aubrecht for the help that he has given me all along my studies in Physics Education.